\def\order#1{{\cal O}\left(#1\right)}
\def\psla{p\hspace{-0.45em}/}
\def\xisla{\xi\hspace{-0.5em}/}
\def\ba{\begin{eqnarray}}
\def\ea{\end{eqnarray}}

\def\pPs{\mbox{p-Ps}}
\def\oPs{\mbox{o-Ps}}

\documentclass{appolb}
\usepackage{psfig}

\begin{document}
\pagestyle{plain}

\newcount\eLiNe\eLiNe=\inputlineno\advance\eLiNe by -1
\title{Positronium properties%
\thanks{Talk given at the 39th
Cracow School of Theoretical Physics, Zakopane, May 1999.
}}
\author{Andrzej Czarnecki
\address{Physics Department\\
Brookhaven National Laboratory\\  Upton, New York 11973}}
\maketitle

\vspace*{-75mm}

\begin{flushright}
BNL-HET-99/38 \\
hep-ph/9911455\\
November 1999
\end{flushright}

\vspace*{60mm}

\begin{abstract}
This talk gives an elementary introduction to the basic properties of
positronium.  Recent progress in theoretical studies of the hyperfine
splitting and lifetime of the ground state is reviewed.  Sensitivity
of these precisely measured quantities to some New Physics effects is
discussed.
\end{abstract}

\section{Introduction}
Positronium, an electron-positron bound state, is a particularly
simple system which offers unique opportunities for testing our
understanding of bound-states in the framework of Quantum
Electrodynamics (QED).  Because its constituents are much lighter (by
a factor of about 200) than any other known charged particles (muons
or pions), the spectrum and lifetimes of positronium states can be
understood with very high precision within an effective theory of
electrons and photons.  Comparison of theoretical predictions
with experiments constrains a variety of New Physics phenomena, such
as axions, millicharged particles, paraphotons, etc.  In addition, the
opportunity of testing the predictions with high accuracy stimulates
development of theoretical tools which can also be applied in other
areas of physics, such as QCD.

This talk gives an elementary introduction to basic properties of
positronium, reviews some recent improvements in their theoretical
description, and briefly touches upon implications for New Physics
searches.

\section{Positronium spectrum}
\label{sec:spectrum}
Positronium (Ps) is an atom resembling  hydrogen in many respects.  To
first approximation its mass differs from the sum of its
constituents' masses ($2m_e\simeq 1.02$ MeV) by the small binding energy,
\ba
E_B = -{m_e \alpha^2 \over 4} \simeq - 6.8 \, {\rm eV},
\ea
which is about half that of hydrogen, because the reduced mass in
positronium is $m_e/2$.  The ground state and its radial excitations
form the so-called gross spectrum, $E_n = E_B/n^2$ ($n=1\ldots$).

This picture, based on the non-relativistic Schr\"odinger equation, has
to be corrected for relativistic effects.  In this talk we will be
concerned entirely with the lowest radial state with zero angular
momentum ($S$ wave).  It comes in two varieties, depending on the sum
of spins of the electron and positron: para-positronium (p-Ps) with
total spin 0, and ortho-positronium (o-Ps), which is a triplet with
total spin 1.  p-Ps is slightly lighter, due to the spin-spin
interaction.  This difference of masses, called the hyperfine
splitting (HFS), offers one of the most accurate tests of
bound-state theory based on QED.  (For a review of other precisely
measured energy intervals in Ps see \eg
\cite{PhK,Czarnecki:1999mw}.)

\begin{figure}[htb]
\hspace*{-0mm}
\psfig{figure=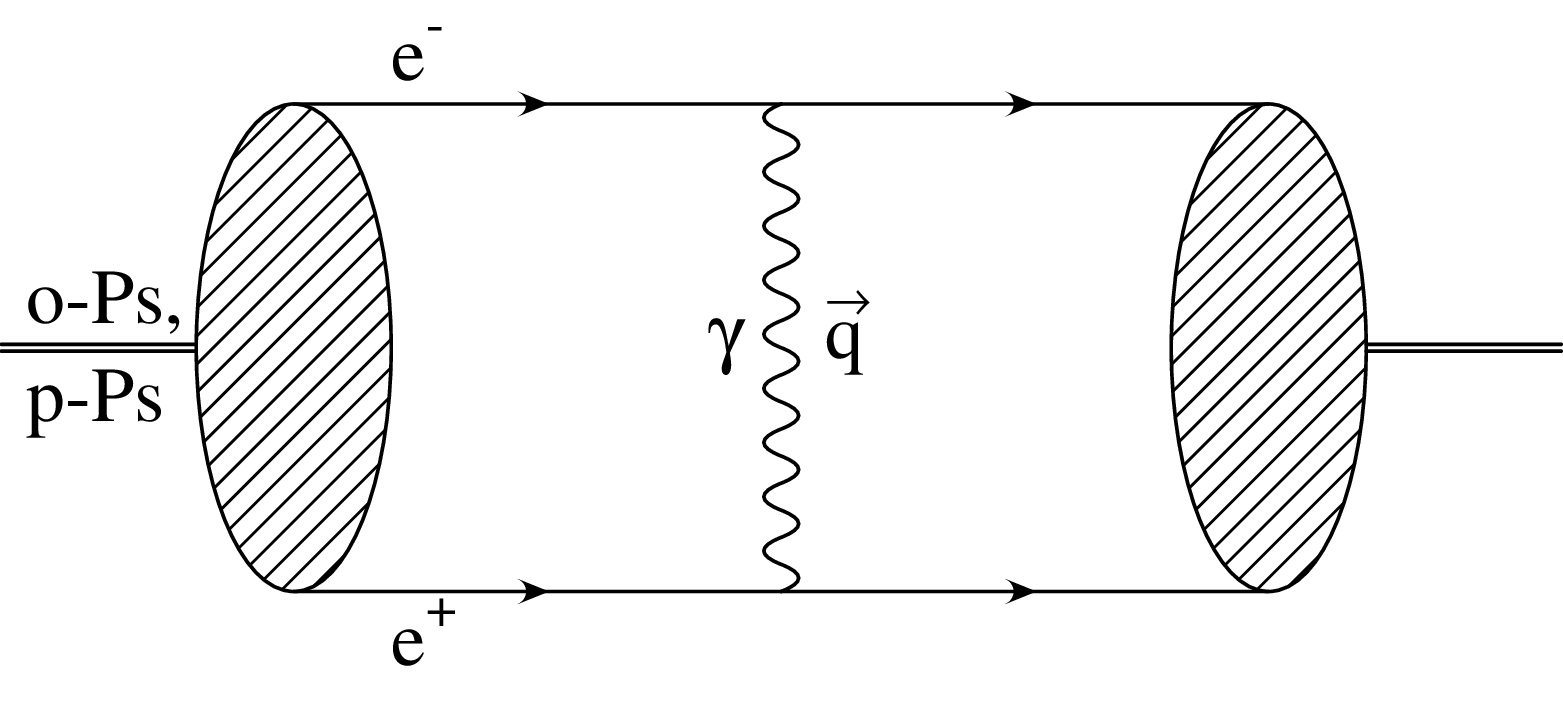,width=55mm}
\hspace*{10mm}
\psfig{figure=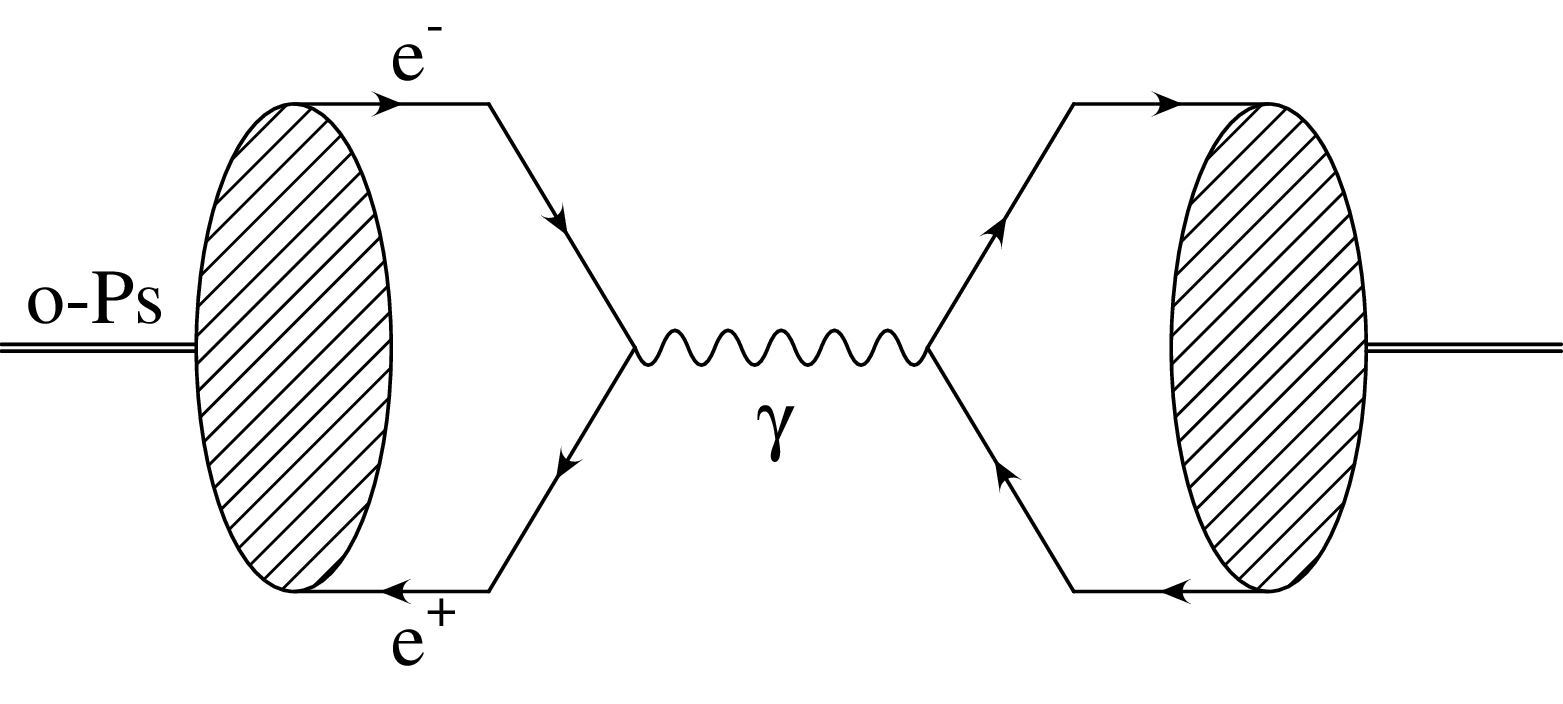,width=55mm}
\\[1mm]
\hspace*{27mm}(a)
\hspace*{60mm}(b)
\\[-2mm]
\caption{Lowest order contributions to HFS in positronium.}
\label{fig:lowestHFS}
\end{figure}

The origin of HFS can be best explained with the lowest-order diagrams
contributing to this effect, shown in Fig.~\ref{fig:lowestHFS}.  They
represent two types of interactions: direct and annihilation.  These
two effects together give the following difference of energy levels of
o-Ps and p-Ps (That difference is conventionally expressed in terms
of the corresponding frequency $\Delta\nu = \Delta E/2\pi \hbar$.  In
this paper we will often use $\Delta\nu$ to denote the energy
intervals):
\ba
\Delta \nu^{(0)} = (\Delta\nu)^D + (\Delta\nu)^A.
\ea
The direct interaction diagram \ref{fig:lowestHFS}(a) represents a
magnetic photon exchange
which induces a spin-dependent potential
\ba
V_M = -{\pi\alpha\over 4m_e^2\vec q^2}
[\sigma_i,\vec \sigma \cdot \vec q]
[\sigma_i',\vec \sigma' \cdot \vec q],
\ea
where un-primed and primed $\sigma$-matrices act on the electron and
positron spinors, respectively.  The difference of expectation values
of this potential in o-Ps and p-Ps states gives
\ba
(\Delta \nu)^D = {m_e\alpha^4 \over 3}.
\ea
The annihilation contribution (Fig.~\ref{fig:lowestHFS}(b)) shifts
only the o-Ps energy and we can evaluate it directly from the
product of amplitudes of o-Ps becoming a photon and of the reverse
process.  Projection on the triplet state can be obtained by taking the
trace with the o-Ps spin wave function, $ {1+\gamma_0\over 2\sqrt{2}}
\xisla $, where $\xi$ is the o-Ps polarization
vector ($\xi^2 = -1$):
\ba
(\Delta \nu)^A &=& -{4\pi\alpha\over (2m_e)^2} |\psi(0)|^2
\left(
\Tr  {1+\gamma_0\over 2\sqrt{2}}\xisla
{\psla - m_e\over 2m_e}
\gamma^\mu
{\psla + m_e\over 2m_e}
\right)^2
\nonumber \\
&=& {m_e\alpha^4\over 4}.
\ea
$|\psi(0)|^2 =  {m_e^3\alpha^3 \over 8\pi}$ is the square of the
Ps wave function at the origin and $p$ denotes the electron (or positron)
four-momentum, in which we neglect the spatial components at this
level of accuracy ($p.\xi=0$).

Adding the two contributions, we find the lowest-order HFS:
\ba
\Delta \nu^{(0)} = (\Delta\nu)^D + (\Delta\nu)^A
 =  {7m_e\alpha^4\over 12}
\simeq 204\,387 \; {\rm MHz}.
\label{eq:zeroHFS}
\ea
This result was obtained in \cite{Pir,Ber,Fer}.  Present experimental
accuracy requires computing one- and two-loop corrections to this
formula.

\begin{figure}[htb]
\hspace*{-0mm}
\psfig{figure=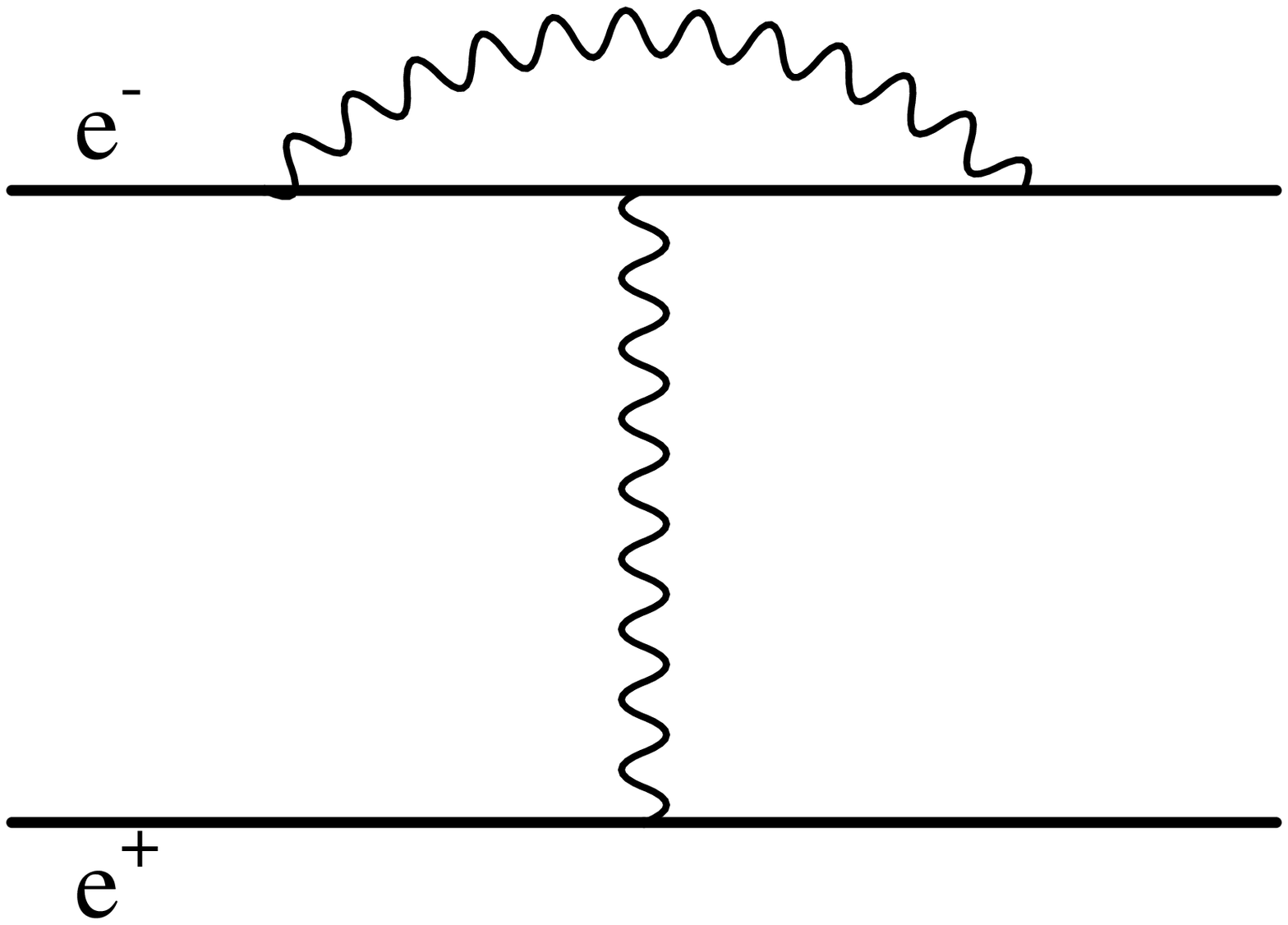,width=35mm}
\hspace*{5mm}
\psfig{figure=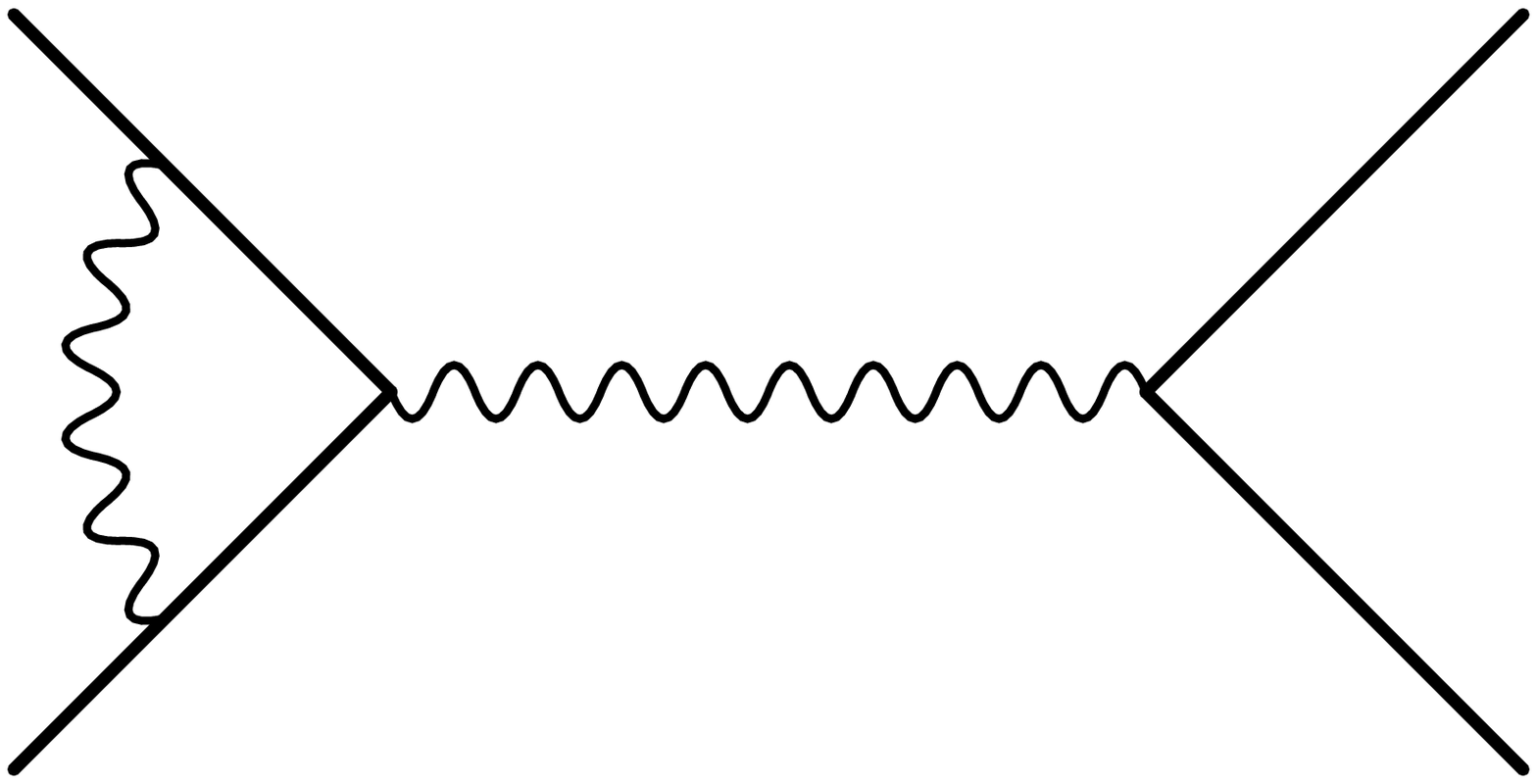,width=35mm,bbllx=72pt,bblly=230pt,%
bburx=540pt,bbury=460pt}
\hspace*{5mm}
\psfig{figure=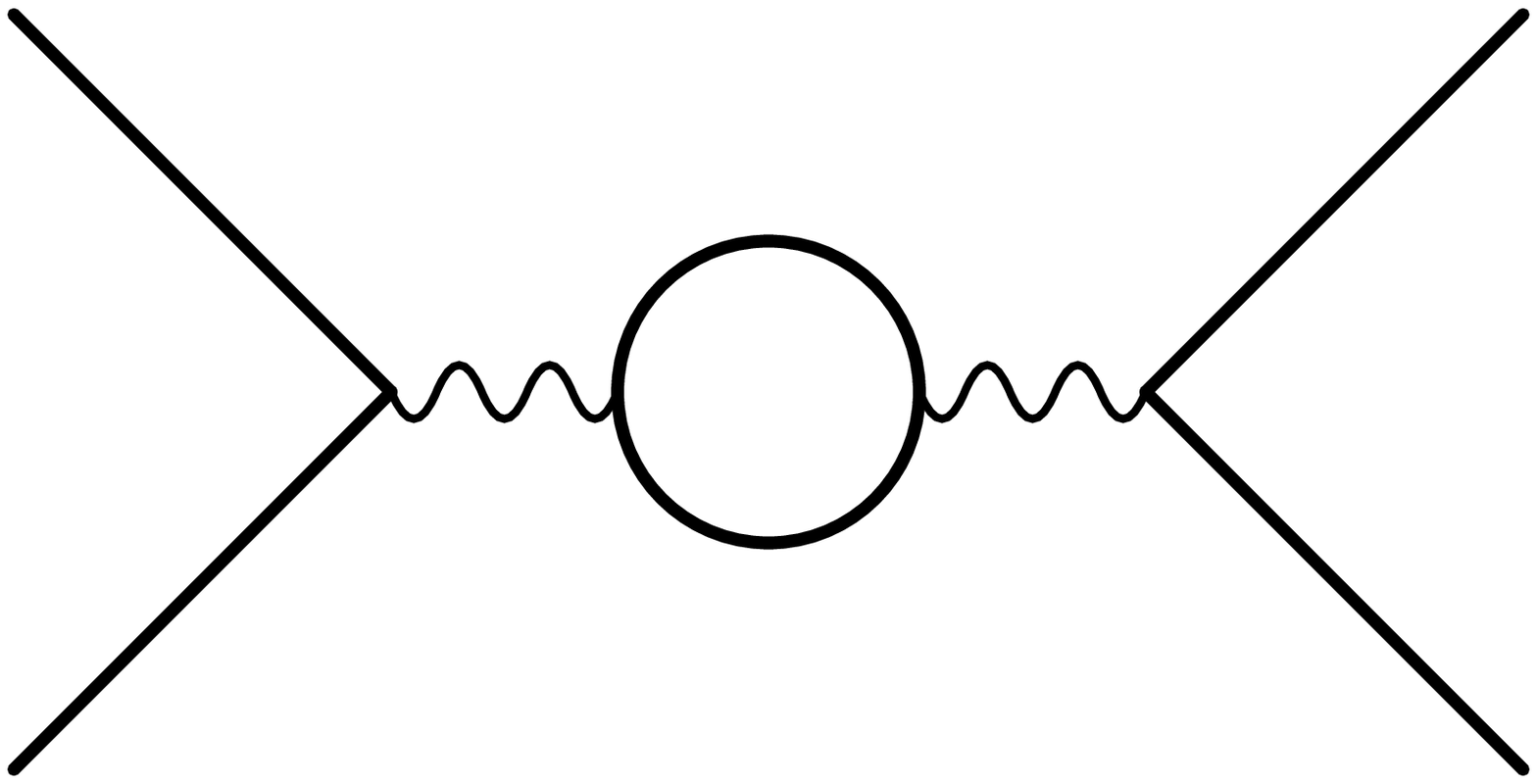,width=35mm,bbllx=72pt,bblly=230pt,%
bburx=540pt,bbury=460pt}
\\[0mm]
\hspace*{17mm}(a)
\hspace*{36mm}(b)
\hspace*{36mm}(c)
\\[5mm]
\psfig{figure=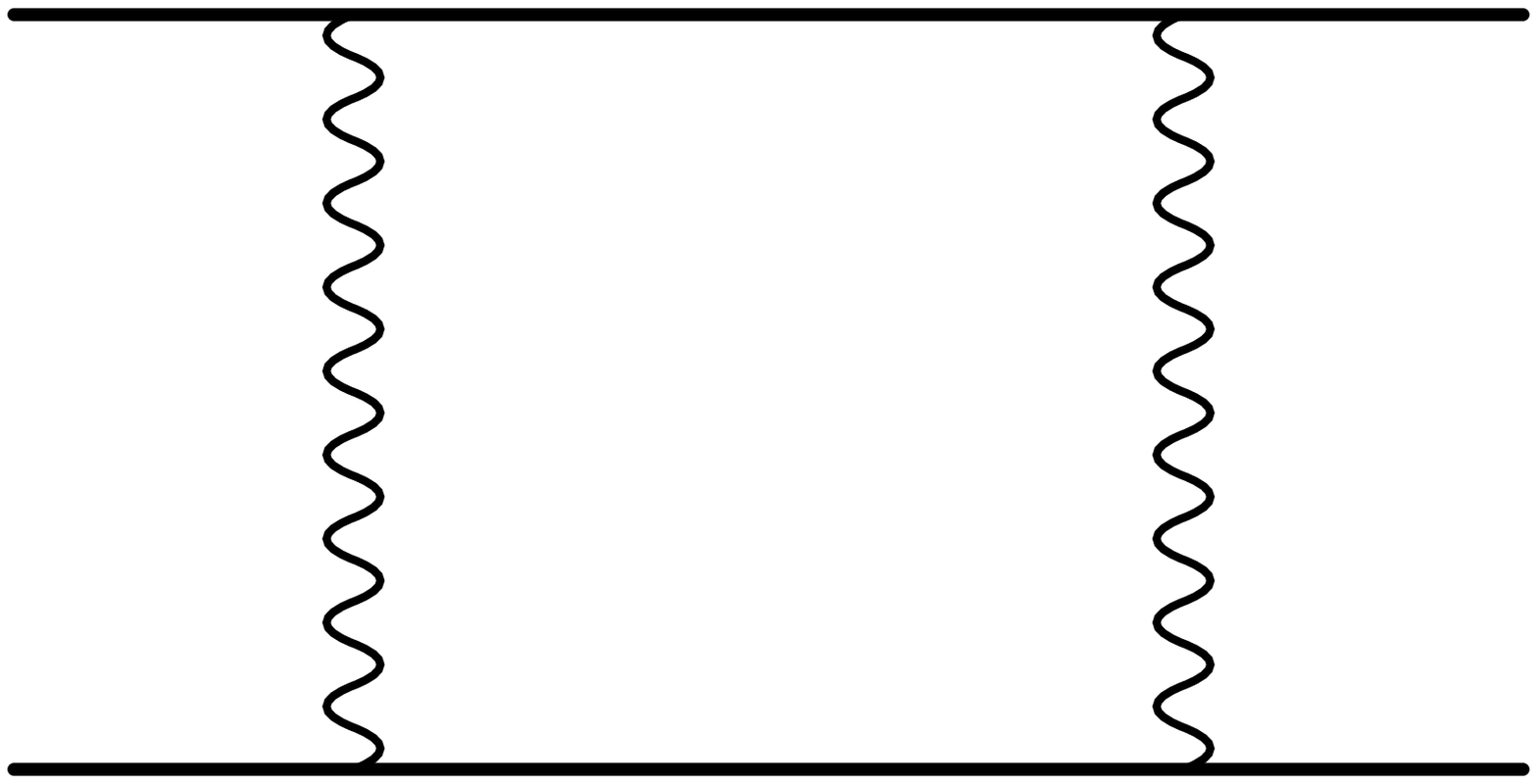,width=35mm}
\raisebox{9mm}{ + \parbox{20mm}{crossed\\photons}}
\hspace*{3mm}
\psfig{figure=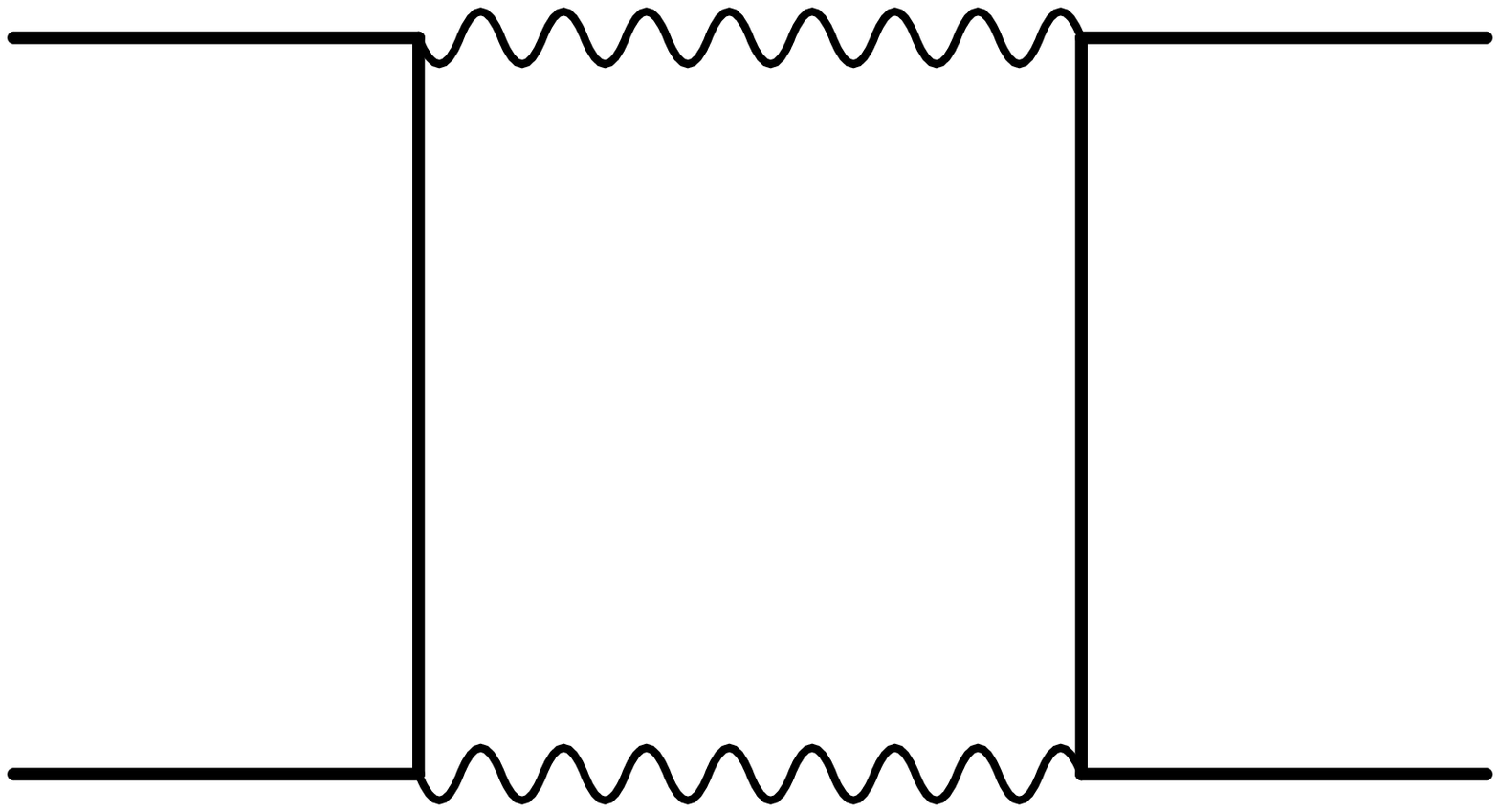,width=35mm}
\raisebox{9mm}{ + \parbox{18mm}{crossed\\photons}}
\\[3mm]
\hspace*{27mm}(d)
\hspace*{55mm}(e)
\\[0mm]
\caption{One-loop corrections to the positronium HFS.}
\label{fig:oneLoopHFS}
\end{figure}
\begin{figure}[htb]
\hspace*{-2mm}
\begin{minipage}{16.cm}
\begin{tabular}{ccc}
\psfig{figure=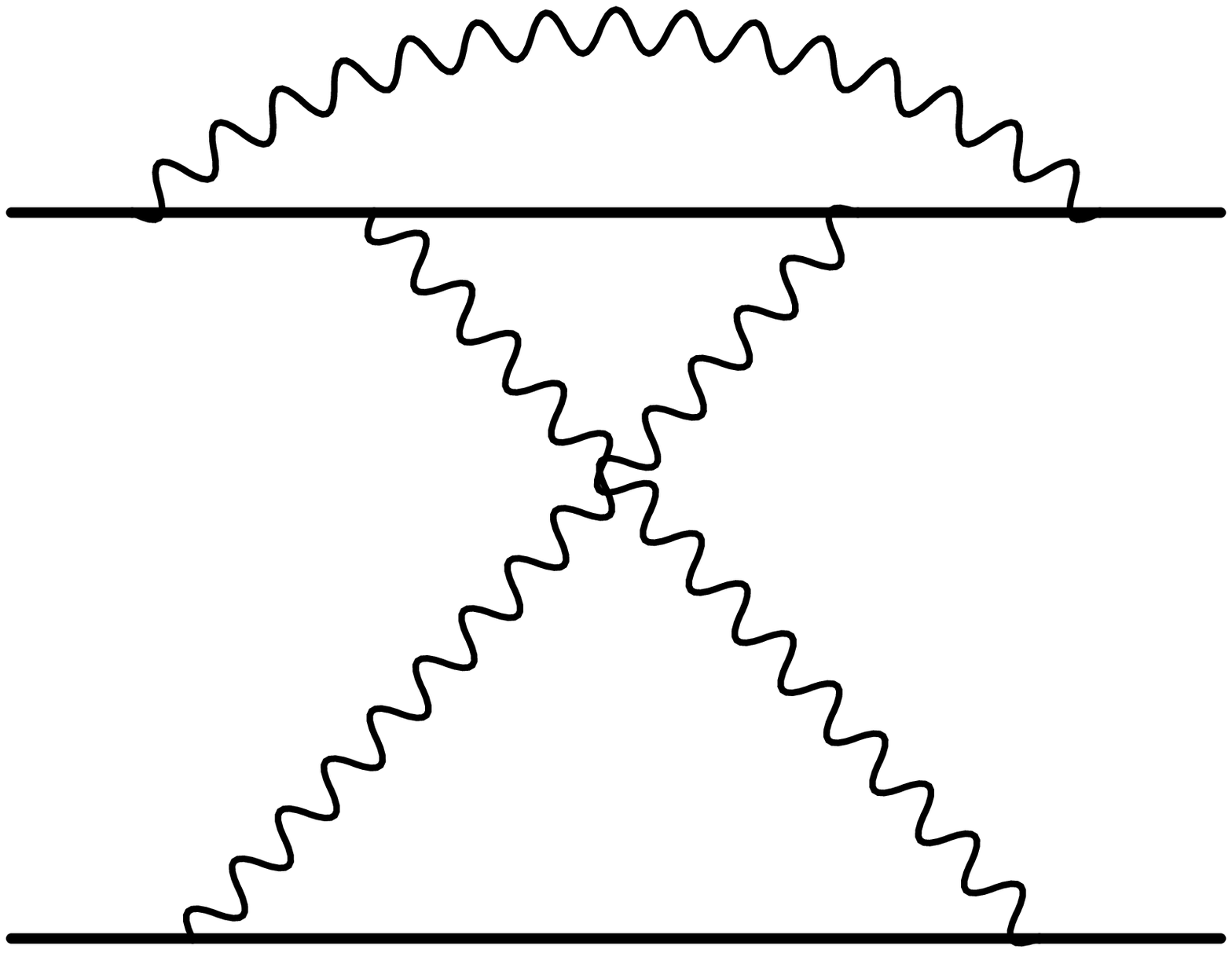,width=33mm} &\hspace{3mm}
\psfig{figure=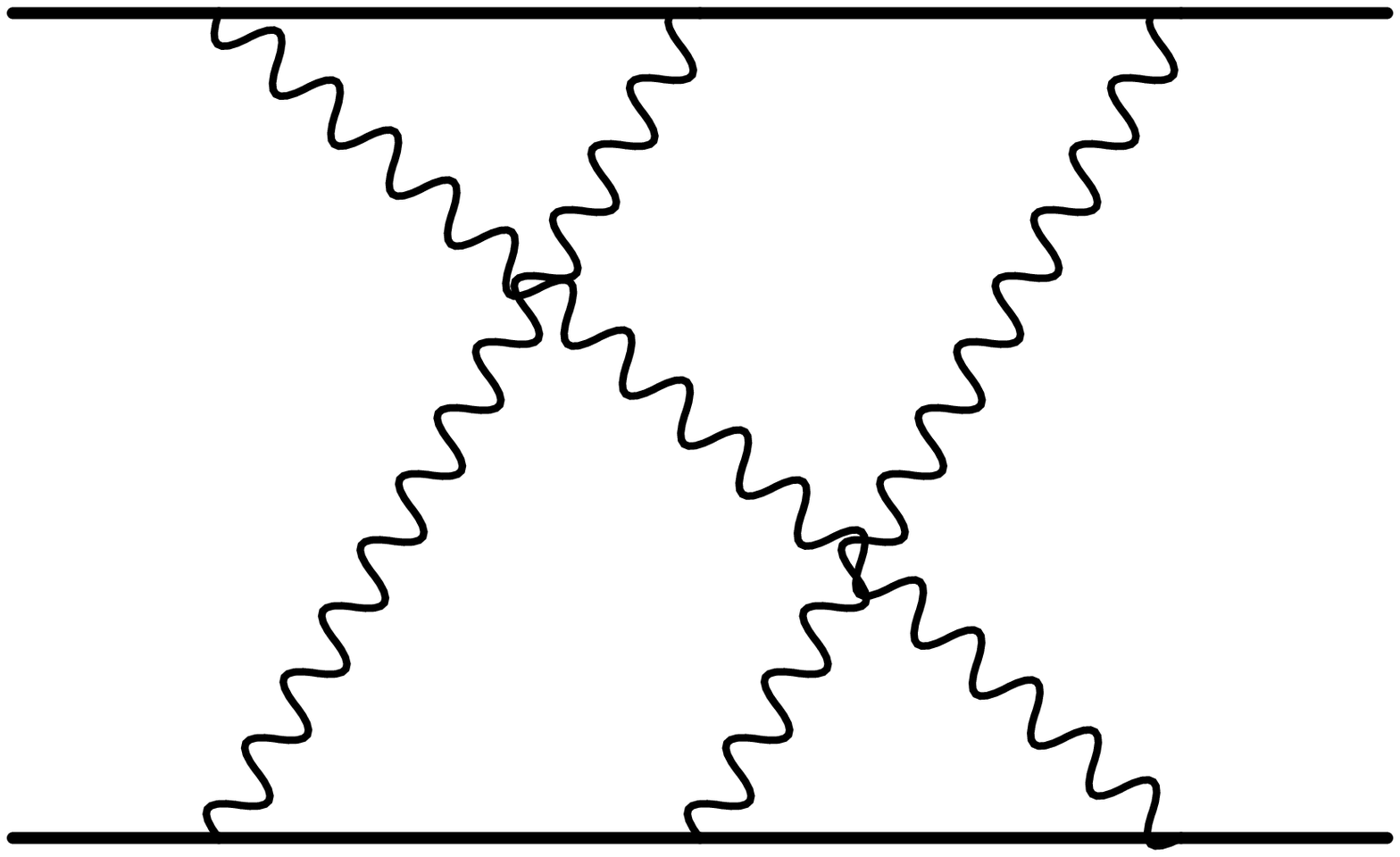,width=33mm} &\hspace{3mm}
\psfig{figure=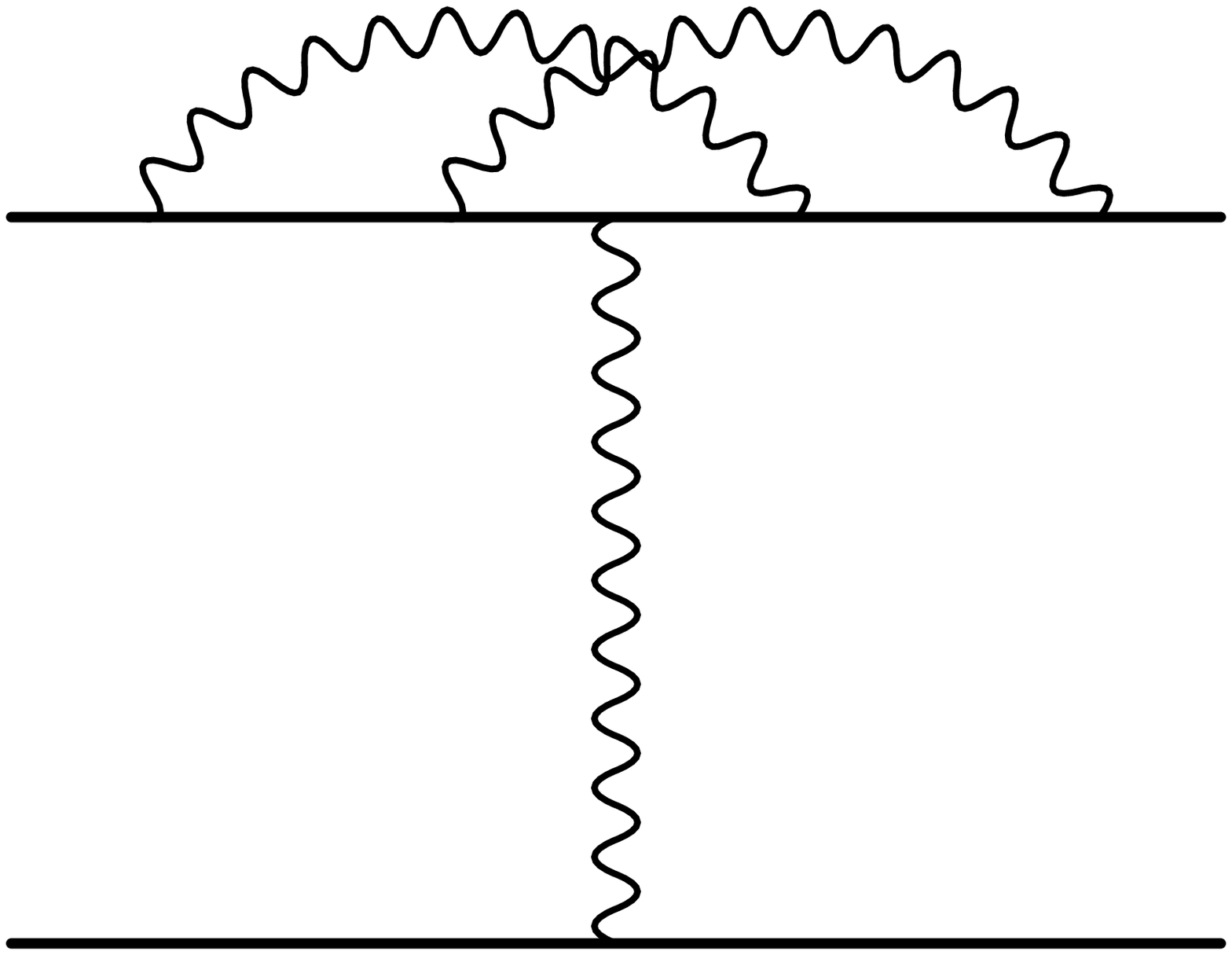,width=33mm}
\\[1.5mm]
(a) & \hspace{2mm} (b) &\hspace{2mm} (c)\\[4mm]
\psfig{figure=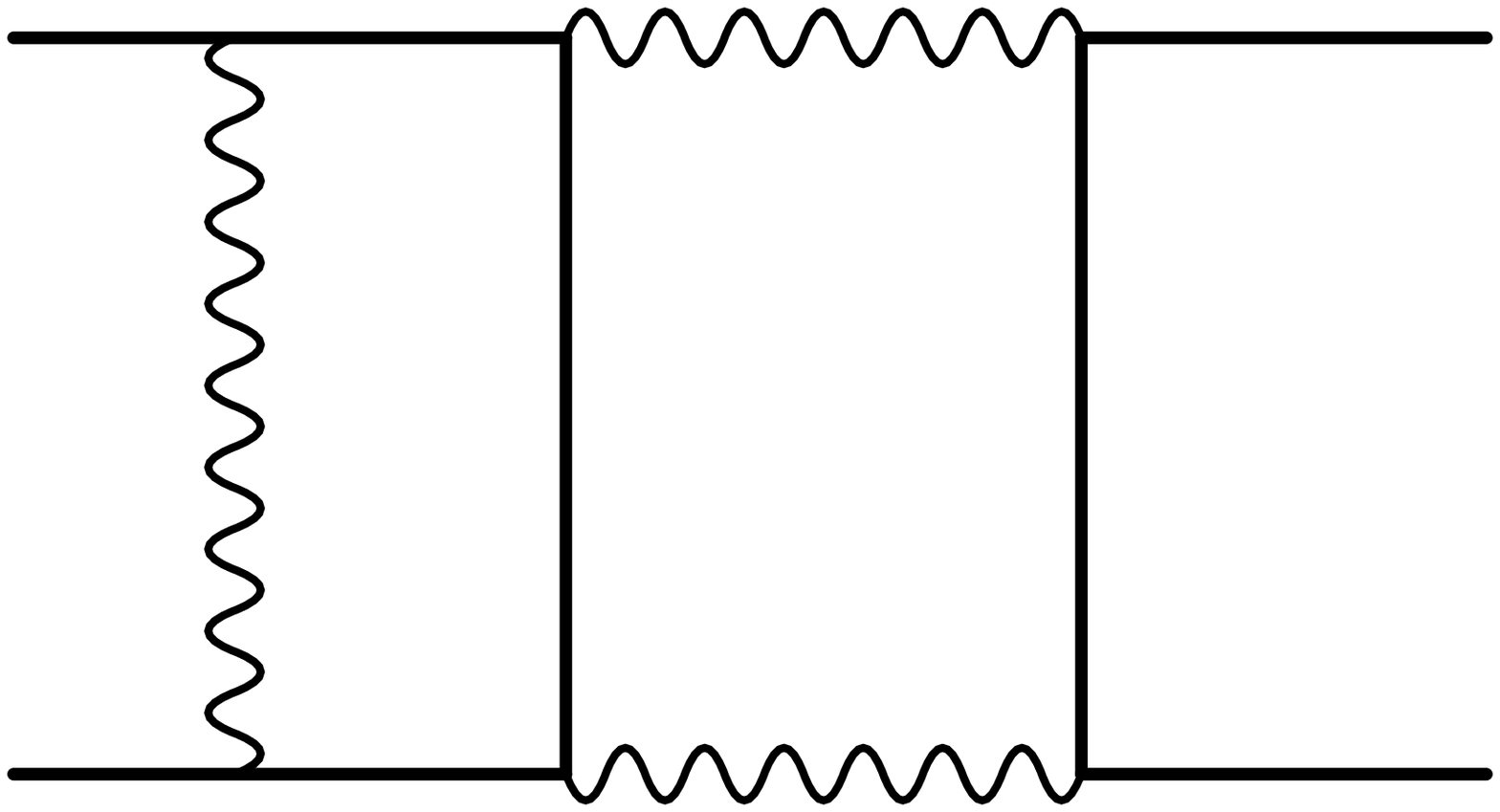,width=33mm} &\hspace{3mm}
\psfig{figure=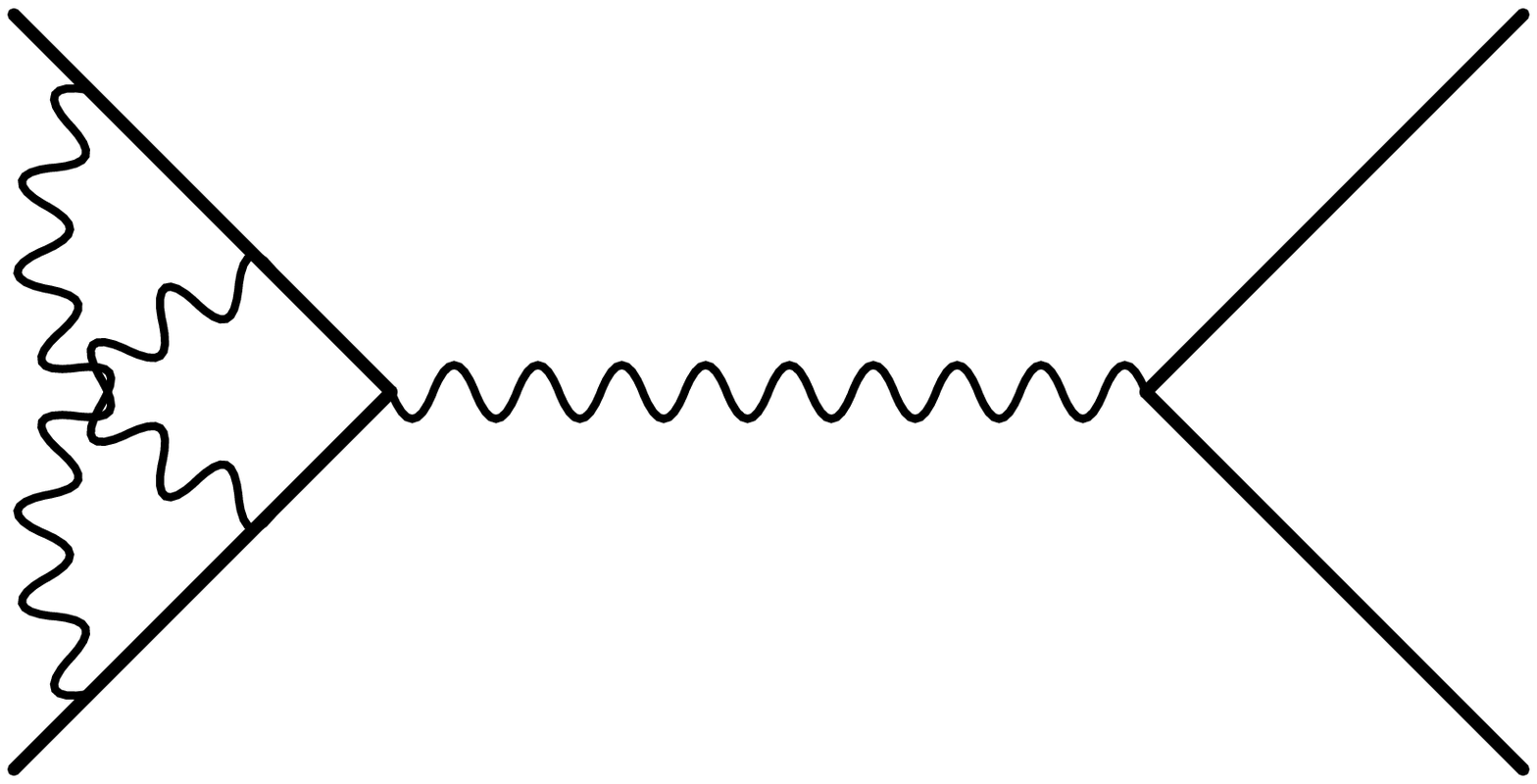,width=33mm} &\hspace{3mm}
\psfig{figure=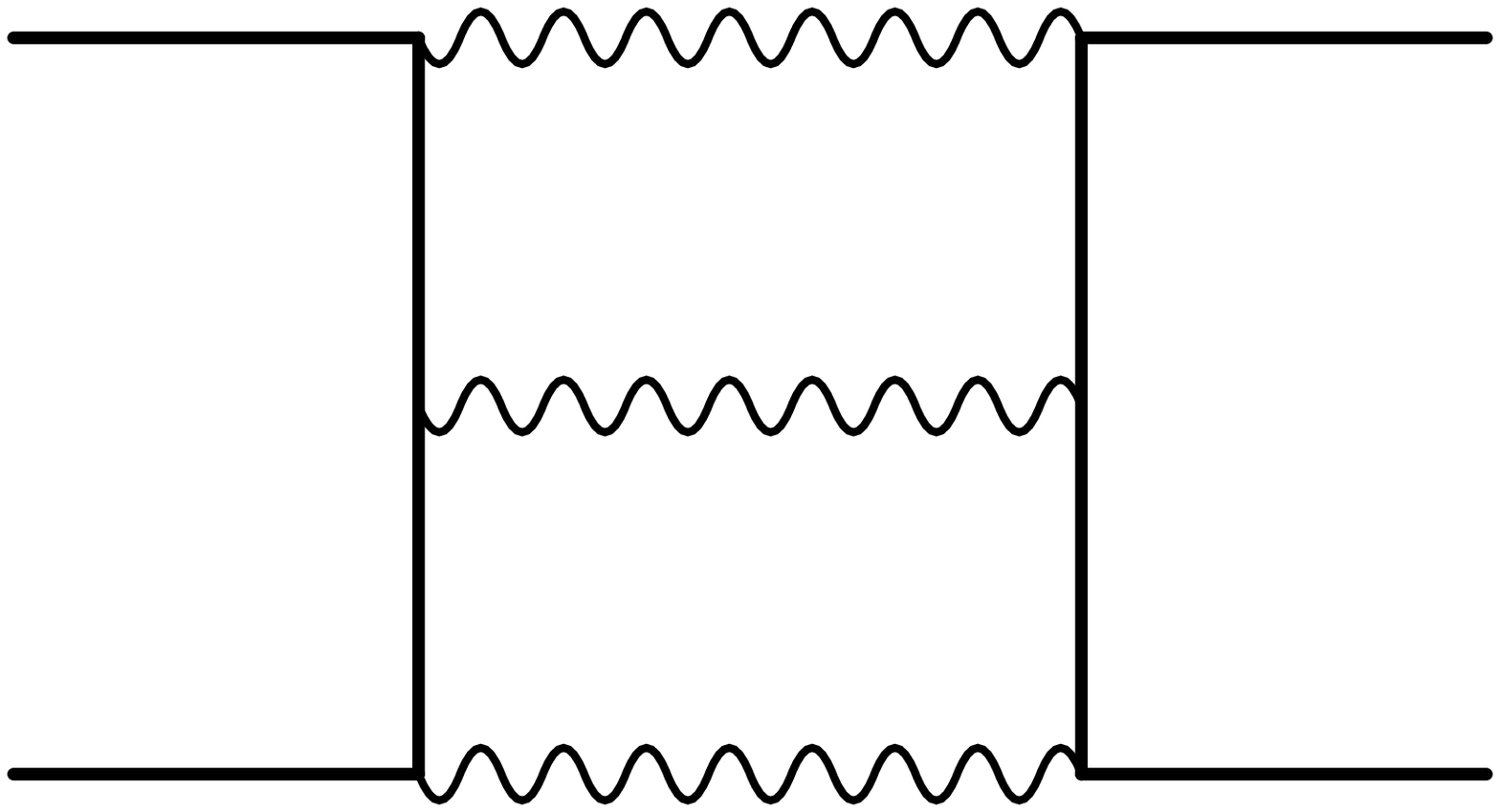,width=33mm}
\\[1.5mm]
(d) & \hspace{2mm} (e) &\hspace{2mm} (f)\\[2mm]
\end{tabular}
\end{minipage}
\caption{Examples of two-loop contributions to the positronium HFS:
(a) radiative-recoil, (b) pure recoil, (c) non-recoil, (d) correction
from the virtual p-Ps annihilation, (e,f) the same with the o-Ps.}
\label{fig:twoLoopHFS}
\end{figure}

The one-loop diagrams are shown in Fig.~\ref{fig:oneLoopHFS}.  They
were calculated by Karplus and Klein \cite{KK}, and found to give the
following contributions to the HFS:
\ba
2a+2d &=& -{1\over 6}{m_e\alpha^5\over \pi},
\nonumber \\
2b &=& -{m_e\alpha^5\over \pi},
\nonumber \\
2c &=&  -{2\over 9}{m_e\alpha^5\over \pi},
\nonumber \\
2e &=&  {1-\ln 2\over 2}{m_e\alpha^5\over \pi}.
\ea
Together with the lowest order result (\ref{eq:zeroHFS}),
they lead to the following corrected formula for the HFS:
\ba
\Delta \nu^{(0)+(1)} =m_e\alpha^4\left[ {7\over 12}-\left({8\over 9}
+{\ln 2\over 2}\right) {\alpha\over \pi}\right].
\ea

At the two-loop level the number of diagrams and their complexity
increase dramatically.  In fact, their analytical 
evaluation was  completed only this year.  Examples of various
types of effects are shown in Fig.~\ref{fig:twoLoopHFS}.
Their evaluation took (with breaks) more than 40 years of efforts by
many groups (see \eg \cite{Czarnecki:1998zv} for references).
Particularly unclear was the status of recoil effects, pictured
in Fig.~\ref{fig:twoLoopHFS}(b).  Until recently there were 3
disagreeing numerical evaluations
\cite{Caswell:1986ui,Ph,AS}.  The difference between the extreme
results was about 5.7 MHz, almost 8 times larger than the present
accuracy of the HFS measurement \cite{Ritter}.

In view of that discrepancy, we recomputed the two-loop recoil
corrections \cite{Czarnecki:1998zv,Czarnecki:1999mw}, using
Non-Relativistic QED (NRQED) \cite{Caswell:1986ui}.  In contrast to
the earlier studies,  dimensional regularization was employed.  In this
way we were able to completely separate the different characteristic energy
scales in the problem and, for the first time,  find an analytical
result for the recoil effects.  Together with the analytical results
obtained earlier  for the remaining radiative-recoil, non-recoil, and
various annihilation contributions, one obtains the following formula
for the positronium HFS, including complete $\order{m_e\alpha^6}$ and
leading-logarithmic \cite{DL} $\order{m_e\alpha^7\ln^2\alpha}$ effects:
\ba
\Delta \nu
&=& m_e\alpha^4 \left \{ \frac {7}{12} -
\frac{\alpha}{\pi} \left (\frac {8}{9}+\frac{\ln 2}{2}  \right )
\right.
\nonumber\\
&&
+\frac {\alpha^2}{\pi^2} \left [
- \frac{5}{24}\pi^2 \ln \alpha
+ \frac {1367}{648}-\frac{5197}{3456}\pi^2
 + \left ( \frac{221}{144}\pi^2 +\frac {1}{2} \right ) \ln 2
-\frac{53}{32}\zeta (3) \right ]
\nonumber\\
&& \left.
-\frac {7\alpha^3}{8\pi}\ln^2\alpha
+\order{\alpha^3\ln\alpha}
 \right \}
\nonumber \\
&=& 203\, 392.01(46) \mbox{ MHz}.
\label{hfsfin}
\ea
The theoretical error quoted has been estimated by taking half
of the last calculated term, the leading-logarithmic
$\order{m_e\alpha^7\ln^2\alpha}$ contribution.  The theoretical
prediction in (\ref{hfsfin}) 
differs by $2.91(74)(46)$ MHz from the most accurate
measurement \cite{Ritter}, where the first error is experimental and
the second theoretical.  If  these errors are combined in quadrature,
this corresponds to a $3.3\sigma$ deviation.  It would be very
interesting to compute further corrections, especially the
next-to-leading $\order{m_e\alpha^7\ln\alpha}$ effects and, even more
important, to re-measure the HFS splitting.

\section{Positronium decays}
In addition to various energy intervals in the Ps spectrum, other
quantities which can be measured with high precision are the lifetimes
of the singlet and triplet ground-states, p-Ps and o-Ps (because of
the possibility of $e^+e^-$ annihilation, both states are unstable) 
\cite{CzKar}.
p-Ps has total spin 0 and can decay into two photons, with a short
lifetime of about $0.125\times 10^{-3}\,\mu$s.  On the other hand,
o-Ps must decay into at least three photons, because a spin 1 state
cannot decay into two photons.  Its lifetime, $\simeq 0.14\,\mu$s, is
about 1000 times longer than that of p-Ps, and is somewhat easier to
accurately determine.

\subsection{Parapositronium decays}

\begin{figure}[htb]
\hspace*{17mm}
\begin{minipage}{16.cm}
\vspace*{3mm}
\begin{tabular}{cc}
\psfig{figure=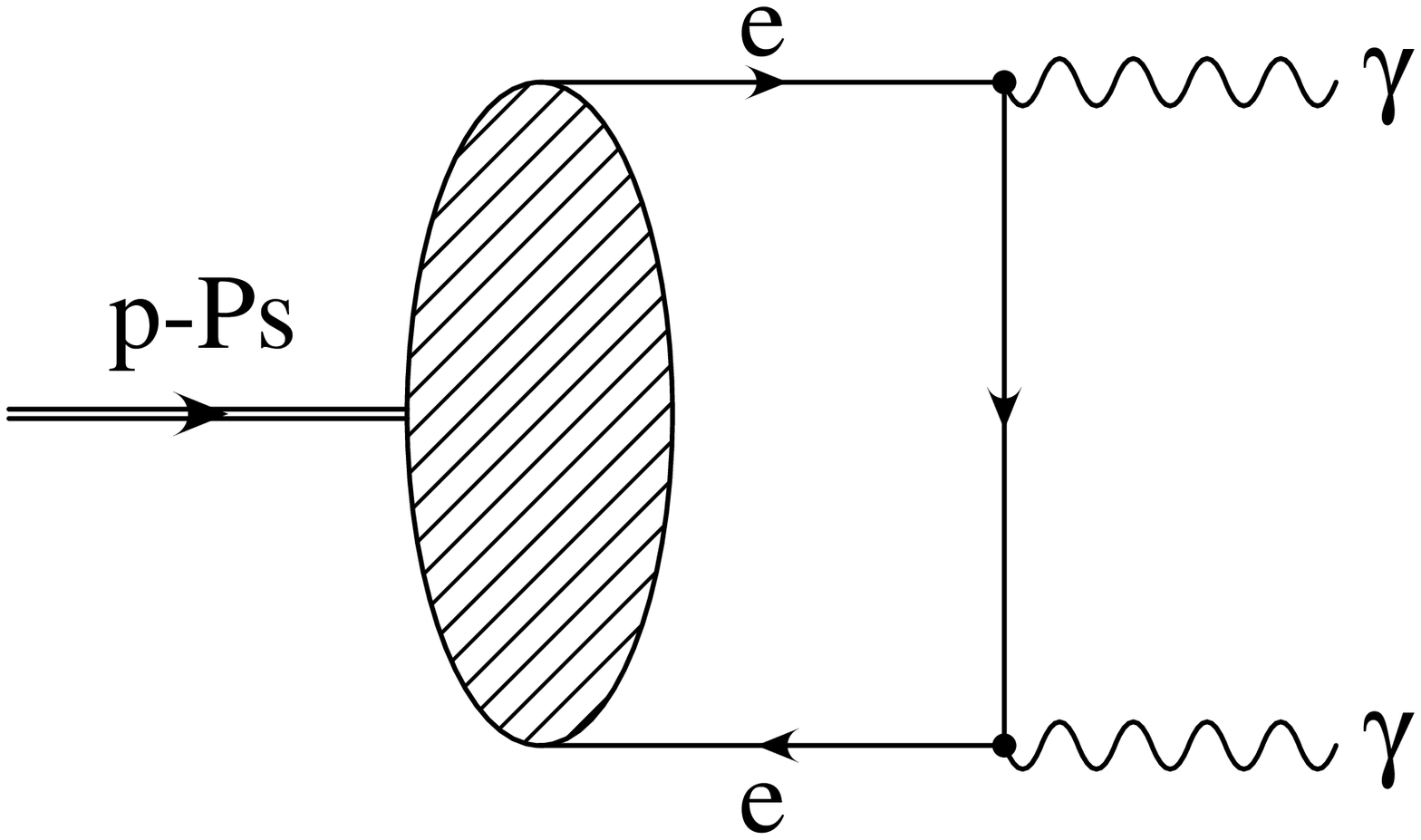,width=35mm,bbllx=72pt,bblly=291pt,%
bburx=544pt,bbury=540pt}
& \hspace*{8mm}
\psfig{figure=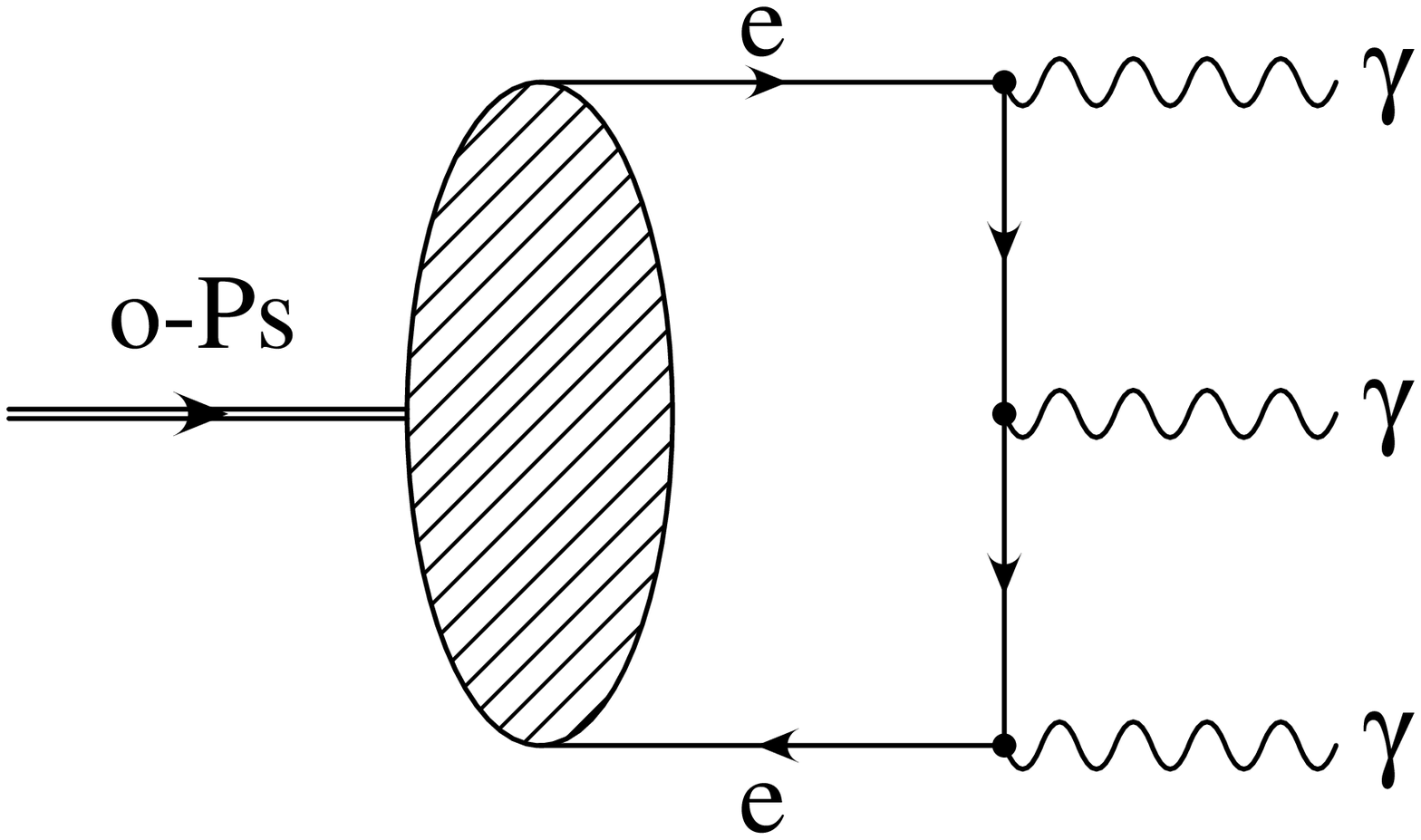,width=35mm,bbllx=72pt,bblly=291pt,%
bburx=544pt,bbury=540pt}
\\[4mm]
\end{tabular}
\end{minipage}
\caption{Lowest order decay channels of p-Ps and o-Ps.}
\label{fig1}
\end{figure}

Barring $C$-violating effects (\eg caused by the weak interactions),
p-Ps can annihilate into only even number of photons (see Fig.~\ref{fig1}).

The decay rate of the p-Ps ground state, $1^1S_0$, can be calculated
as a series in $\alpha$.  The two-photon decay rate is
\ba
\Gamma(\mbox{p-Ps}\to \gamma\gamma) &=&
{m_e\alpha^5\over 2}
\left[ 1-\left(5-{\pi^2\over 4}\right) {\alpha\over \pi}
+2\alpha^2 \ln {1\over \alpha}
+ 1.75(30) \left( {\alpha\over\pi}\right)^2
\right.
\nonumber \\
&& \left.
- {3\alpha^3 \over 2\pi} \ln^2 {1\over \alpha}
+\order{\alpha^3\ln {1\over \alpha} } \right]
= 7989.50(2)~\mu{\rm s}^{-1},
\label{eq:pps}
\ea
where the non-logarithmic terms $\order{\alpha^2}$
\cite{Czarnecki:1999gv,Czarnecki:1999ci} and leading-logarithmic terms
$\order{\alpha^3\ln^2\alpha}$ \cite{DL} have been obtained only
recently.

The four--photon branching ratio is of relative order $\alpha^2$
\cite{Billoire:1978wq,Muta:1982hb,AdkBr}:
\begin{equation}
\label{B24}
{\rm BR}(\mbox{p-Ps}\to 4\gamma)={\Gamma(\mbox{p-Ps}\to \gamma\gamma)
\over \Gamma(\mbox{p-Ps}\to 4\gamma)} =
0.277(1)\left(\frac{\alpha}{\pi}\right)^2
\simeq 1.49\cdot 10^{-6}\,.
\end{equation}
The theoretical prediction, (\ref{eq:pps}), agrees well
with the experiment \cite{AlRam94},
\begin{equation}
\Gamma_{\rm exp}(\pPs)= 7990.9(1.7)~\mu{\rm s}^{-1}.
\end{equation}

\subsection{Orthopositronium decays}
The ground state of  orthopositronium, $1^3S_1$,
can decay into an odd number of the photons only (if $C$ is conserved).
The three--photon (see Fig.~\ref{fig1}) decay 
rate is given by
\ba
\Gamma(\mbox{o-Ps}\to \gamma\gamma\gamma)&=&
{2(\pi^2-9)m_e\alpha^6\over 9\pi}
\left[ 1-10.28661{\alpha\over \pi}
-{\alpha^2\over 3}\ln {1\over \alpha}
+ B_o \left( {\alpha\over\pi}\right)^2
\right.
\nonumber \\
&&
\left.
\hspace*{-26mm}
- {3\alpha^3 \over 2\pi} \ln^2 {1\over \alpha}
+\order{\alpha^3\ln \alpha}\right]
\simeq
\left(
 7.0382  +   0.39 \cdot 10^{-4} \; B_o
\right)
\,\mu {\rm s}^{-1}.
\label{eq:otheor}
\ea
Because of its three-body phase space and a large number of diagrams,
a complete theoretical analysis of o-Ps decays is more difficult than in the
case of p-Ps.  The non-logarithmic two-loop effects, parameterized by
$B_o$, have not been evaluated as yet, except for a subset of the so-called
soft corrections.  Those partial results depend on the scheme adopted
for regularizing ultraviolet divergences and do not give a reliable
estimate of the complete $B_o$.  Further theoretical work is needed to
find that potentially important correction.

Five-photon decay branching ratio is of order $\alpha^2$
\cite{AdkBr,Lepage:1983yy},
\begin{equation}
{\rm BR}(\mbox{o-Ps}\to 5\gamma)={\Gamma(\mbox{o-Ps}\to 5\gamma)
\over \Gamma(\mbox{o-Ps}\to  \gamma\gamma\gamma)}
= 0.19(1)\left({\alpha\over \pi}\right)^2
\simeq 1.0\times  10^{-6},
\label{B35}
\end{equation}
and does not significantly influence the total width.

Table~\ref{Tortho} lists the three latest experimental
results for the o-Ps lifetime.
\begin{table}[h]
\caption{Recent experimental results for the o-Ps lifetime.  ``Method''
in the second column refers to the medium in which o-Ps decays.  The
last column shows the value of the two-loop coefficient $B_o$,
necessary to bring the theoretical prediction
(\protect\ref{eq:otheor}) into agreement with the
given experimental value.  The last line gives the theoretical
prediction with $B_o=0$.}
\label{Tortho}
\begin{center}
\begin{tabular}{l @{\hspace*{10mm}}c @{\hspace*{10mm}}l @{\hspace*{10mm}}r}
\hline \hline
&&&\\
Reference & Method & $\Gamma(\mbox{o-Ps})~[\mu s^{-1}]$ & $B_o$ \\
&&&\\
\hline
&&&\\
Ann Arbor \protect{\cite{Westbrook89}}&Gas&7.0514(14)&338(36)\\
&&&\\
Ann Arbor \protect{\cite{Nico:1990gi}} & Vacuum&7.0482(16)&256(41)\\
&&&\\
Tokyo \protect{\cite{Asai:1995re}} & Powder&7.0398(29)&41(74)\\
&&&\\
\hline
\multicolumn{2}{c}{}&&\\
\multicolumn{2}{c}{Theory without $\alpha^2$}& 7.0382&0\\
\multicolumn{2}{c}{}&&\\[-1mm]
\hline
\hline
\end{tabular}
\vspace*{2mm}
\end{center}
\end{table}

The last column indicates the value of the two-loop coefficient $B_o$
necessary to reconcile a given central experimental value with the
theoretical prediction (\ref{eq:otheor}).  We see that the most
precise Ann Arbor experiments require an anomalously large value of
$B_o$.  This has been known as the ``o-Ps lifetime puzzle.''
More recent Tokyo results, which are at present somewhat less
accurate, are in good agreement with the QED prediction with a much
smaller $B_o$.  A new experiment is underway at Tokyo, and the
data collected so far are in agreement \cite{AsaiPriv} with the
previous result \cite{Asai:1995re}.  This ongoing effort is expected to
provide a measurement of $\Gamma(\oPs)$ with an accuracy of about 150
ppm, somewhat better than the presently available Ann Arbor results.

Concerning a possible large value of $B_o$,
it should be mentioned that the perturbative coefficients are moderate
in QED predictions for other observables studied with high accuracy,
such as leptonic anomalous magnetic moments ($g-2$), or various
spectroscopic properties of positronium and muonium.  If there are
exceptions, they are caused by widely separated energy scales, such as
the electron and muon masses in the muon $g-2$.  In the
Ps lifetime such effects arise as logarithms of $\alpha$
and have already been accounted for in the leading order.

It has also been argued that the smallness of the complete two-loop
non-logarithmic effects in p-Ps, which have recently been calculated
\cite{Czarnecki:1999gv,Czarnecki:1999ci}, indicates that analogous
two-loop effects in o-Ps are unlikely to explain the large discrepancy
with experiment.  However, this issue remains controversial and can
only be clarified by an explicit calculation of $B_o$, and renewed
experimental studies of $\Gamma(\mbox{o-Ps})$.

\section{Some implications for New Physics searches}
It is interesting to compare the sensitivity of Ps decays and HFS
measurements to various kinds of New Physics.  Here we focus on the
example of  millicharged particles (for a review see
\cite{Davidson:1991si}).

The millicharged particles we are interested in here are fermions,
similar in properties to the electron, but with an unknown small mass
$m_x$ and electric charge $\eta e$ (where $-e$ is electron's charge
and $\eta\ll 1$).  Such particles were searched for in decays of o-Ps
\cite{Mits93}, where a 90\% confidence level upper bound
$\eta<8.6\times 10^{-6}$ was found for $m_x \ll m_e$ (this bound is
not very sensitive to $m_x$, except for $m_x$ close to $m_e$, where it
becomes weaker).  That bound was found by comparing a theoretical
prediction for the decay rate of o-Ps into a pair $X\bar X$ of
millicharged particles,
\ba
\Gamma(\mbox{o-Ps}\to X\bar X) = \eta^2 {m_e\alpha^6\over 6}
\sqrt{1-{m_x^2\over m_e^2}} \left( 1+{m_x^2\over 2m_e^2}\right),
\ea
with the measured bound on the maximum decay rate of o-Ps into
invisible particles, $\Gamma(\mbox{o-Ps}\to {\rm invisible}) <
2.8\times 10^{-6}\Gamma(\mbox{o-Ps}\to \gamma\gamma\gamma)$.

How does this compare with the sensitivity of HFS to possible extra
loops in the photonic vacuum polarization
({\em cf.}~Fig.~\ref{fig:oneLoopHFS}(c))?  We find that a pair of
millicharged particles would have a negative contribution to Ps HFS,
given by the formula
\ba
\Delta \nu({\rm millicharged}) &=&
{\eta^2 m_e\alpha^5\over 12\pi}
\nonumber \\
&&
\hspace*{-31mm}
\times
\left\{
{1\over 3}
+2\left( 1+{m_x^2\over 2m_e^2}\right)
\left[
\sqrt{\left({m_x^2\over m_e^2}-1\right)}
      {\rm arccot}\sqrt{\left({m_x^2\over m_e^2}-1\right)}
-1
\right]
\right\}
\nonumber \\
&\simeq& -{\eta^2 m_e^3\alpha^5\over 15\pi m_x^2} \qquad
\mbox{(for $m_x\gg m_e$)}.
\ea
We note that the o-Ps decay search for $X\bar X$ has the advantage
that the Standard Model background (o-Ps$\to \gamma\gamma\gamma$) is
suppressed by an additional factor of $\alpha^2$.  Therefore, with
presently achievable experimental accuracy, HFS cannot compete with
o-Ps if $m_x$ is smaller than $m_e$.  For this reason we have
displayed the simple limiting behavior of the HFS contribution for
large $m_x$.

Another potentially interesting observable which might be sensitive to
$X\bar X$ effects is the electron anomalous magnetic moment
($a_e=(g_e-2)/2$).  If $m_x\gg m_e$, the contribution of an $X\bar X$
loop is
\ba
\Delta a_e({\rm millicharged}) = {\eta^2\alpha^2 m_e^2\over 45\pi^2
m_x^2}.
\label{eq:ae}
\ea
We can now express the millicharged pair contribution to the HFS by
its effect on $a_e$:
\ba
\Delta \nu({\rm millicharged})
&=&
-3\pi m_e\alpha^3
\Delta a_e({\rm millicharged})
\nonumber \\
&=&-{36\pi\over 7\alpha} \Delta \nu^{(0)}\Delta a_e({\rm millicharged}),
\ea
where $\Delta \nu^{(0)}$ denotes the lowest order Ps HFS, given in
(\ref{eq:zeroHFS}).  In the lowest order in $\alpha$ we have
$a_e^{(0)} = \alpha/2\pi$, so that
\ba
{ \Delta \nu({\rm millicharged}) \over \Delta \nu^{(0)}}
= -{18\over 7} {\Delta a_e({\rm millicharged}) \over a_e^{(0)}}.
\ea
Since the present agreement of theory and experiment for $a_e$ is at
the level of 3 parts in $10^8$ \cite{Czarnecki:1998nd}, we see that
the largest compatible effect of millicharged particles in HFS could
be of the order of $10^{-7}$.  This is insufficient to reconcile the
theory with experiment since, as we saw in Section \ref{sec:spectrum},
the observed HFS value is smaller than the theoretical prediction by
about 3 MHz, or $1.4\times 10^{-5} \Delta \nu^{(0)}$.  It might,
however, be worth noting that the differences of central theoretical
and experimental values of $a_e$ and Ps HFS have opposite signs, and
an $X\bar X$ pair would decrease both discrepancies.

Positronium decays were also used to search for light pseudoscalar
bosons (see \eg \cite{Asai:1991rd}).  It turns out that such searches
give much more stringent constraints on the coupling of pseudoscalars
to electrons than do present HFS measurements  (by a few orders of
magnitude).  If pseudoscalars are heavier than o-Ps, the relative
sensitivity of $a_e$ and Ps HFS to such particles was discussed in
\cite{Krauss:1986bw}.

\section{Summary}

We have reviewed the basic properties of the positronium spectrum and
decay modes.  We have seen that there are discrepancies between the
theoretical predictions and experimental determinations of the
hyperfine splitting (by 3.3$\sigma$) and o-Ps lifetime (by about
$6-9\sigma$ for Ann Arbor experiments).  It is unlikely that New
Physics effects are responsible for such large differences.  It is
also very unlikely that the uncalculated higher-order effects in QED
could alone account for such discrepancies.  Most likely, both
problems could be clarified by new experiments.  While efforts are
underway to obtain a more accurate value of the o-Ps lifetime, it
would also be very useful to re-measure the hyperfine splitting.

\section*{Acknowledgements}
I am grateful to S. Asai for helpful discussions on o-Ps lifetime
measurements, to S. Davidson and W. Marciano for careful reading of
the manuscript and helpful remarks, and to K. Melnikov and
A. Yelkhovsky for collaboration on positronium physics.  I thank
Professor Andrzej Bia{\l}as and Professor Micha{\l} Prasza{\l}owicz
for inviting me to the Cracow School of Theoretical Physics.  This
research was supported by DOE under grant DE-AC02-98CH10886.


\end{document}